\documentclass[aoas,secthm,nameyear,dvips]{arximspdf}

\doi{10.1214/10-AOAS406}
\volume{5}
\issue{2A}
\pubyear{2011}
\firstpage{669}
\lastpage{683}

\makeatletter
\newcommand{\bQ}{\mathbf{Q}}
\newcommand{\bz}{\mathbf{z}}
\newcommand{\bX}{\mathbf{X}}
\newcommand{\bXi}{\bolds{\Xi}}
\newcommand{\bDelta}{\bolds{\Delta}}

\newtheorem{prop}{Proposition}[section]
\makeatother

\begin{document}
\begin{frontmatter}

\title{Quantum Monte Carlo simulation\thanksref{AUT1}}
\thankstext{AUT1}{Supported in part by the NSF Grant DMS-10-05635.}
\pdftitle{Quantum Monte Carlo simulation}
\runtitle{Quantum simulation}

\begin{aug}
\author[A]{\fnms{Yazhen} \snm{Wang}\corref{}\ead[label=e1]{yzwang@stat.wisc.edu}}
\runauthor{Y. Wang}
\affiliation{University of Wisconsin-Madison}
\address[A]{Department of Statistics \\
University of Wisconsin-Madison\\
1300 University Avenue \\
Madison, Wisconsin 53706 \\
USA \\
\printead{e1}} %adresu isvedimo komanda gale!
\end{aug}

% HISTORY:
\received{\smonth{1} \syear{2010}}
\revised{\smonth{8} \syear{2010}}

% ABSTRACT
%
\begin{abstract}
Contemporary scientific studies often rely on the understanding of
complex quantum systems via
computer simulation. This paper initiates the statistical study
of quantum simulation and proposes a~Monte Carlo method for estimating
analytically intractable quantities. We derive the bias and variance
for the proposed Monte
Carlo quantum simulation estimator and establish the asymptotic theory
for the estimator. The theory is used to design
a computational scheme for minimizing the mean square error of the estimator.
\end{abstract}

% KEYWORDS
%

\end{frontmatter}

%s1 ###
\section{Introduction}\label{sec1}
Computer-aided simulations of physical systems are wi\-dely used in
scientific and
engineering studies such as aircraft and car design and nuclear
explosion modeling.
While the traditional simulation methods with the aid of classical
computers based on
transistors are to understand basic properties of materials, many
contemporary simulations
rely on understanding quantum systems, such as those in bio-chemistry
and nano-technology
for the design of nano-materials and novel molecules.
See Aspuru-Guzik et al. (\citeyear{Aetal2005}), Kou (\citeyear{K2009}) and Waldner (\citeyear{W2007}).

A quantum system is described by its state, which is often
characterized by a vector in
some complex Hilbert space. The number of complex numbers required to
characterize the quantum
state normally grows exponentially with the size of the system, rather
than linearly,
as occurs in classical physical systems. Consequently, for a quantum system
it takes an exponential number of bits of memory on a classical
computer to store its quantum
state, and simulations of quantum systems via classic computers face
great computational challenge.
As quantum systems are able to store and keep track an exponential
number of complex numbers and perform data manipulations and
calculations as the systems evolve,
quantum computation and quantum information are to grapple with
understanding how to take advantage of the enormous
information hidden in quantum systems and to harness the immense
potential computational power of
atoms and molecules for the purpose of information processing and computation.
Quantum computers built upon quantum systems may excel in the
simulation of naturally occurring
quantum systems, where such quantum systems may be hard to simulate in
an efficient manner by
classical computers [Abrams and Lloyd (\citeyear{AL1997}), Boghosian and Taylor
(\citeyear{BT1998}) and Zalka (\citeyear{Z1998})].

To the best of our knowledge, this paper is the first to introduce
quantum computation and study
quantum simulation in the statistical framework. Specifically, we will
propose a Monte Carlo quantum simulation method for computing analytically
intractable quantities and analyze approximation errors and random
variations of the proposed Monte
Carlo estimator. The theoretical analysis establishes a strategy to
design an optimal scheme for utilizing
computational resources in obtaining the Monte Carlo estimator.

The rest of the paper proceeds as follows. Section \ref{sec2} provides a brief
review on quantum mechanics,
quantum statistics and basic concepts of quantum computation. Section \ref{sec3}
proposes a Monte Carlo
quantum simulation method and then presents the statistical analysis
for the method. We derive the variance and
bias for the proposed estimator and establish the strategy to allocate
computational resources in the Monte Carlo quantum simulation for
minimizing the mean square error
of the estimator.
A quantum simulation example is illustrated in Section \ref{sec4}.

%s2 ###
\section{Brief background review}\label{sec2}
%s2.1 ###
\subsection{Quantum physics}\label{sec2.1}
Quantum mechanics describes phenomena at microscopic level such as position
and momentum of an individual particle like an atom or electron, spin
of an
electron, detection of light photons, and the emission and absorption of
light by atoms. Unlike classical mechanics where measurements of quantities
like position and momentum can be observed %very
accurately, the quantum
theory can only make statistical prediction about the results of the
measurements performed.

Mathematically quantum mechanics is usually described by a Hilbert
space $\mathcal{H}$
and Hermitian (or self-adjoint) operators on $\mathcal{H}$. As in quantum mechanics,
we adopt standard Dirac notation $|\cdot\rangle$, which
is called a ket, to indicate that the object is an element in $\mathcal{H}$.
A quantum system is completely
described by its state and the time evolution of the state. A state is often
classified as a pure state or an ensemble of pure states that are easy
to describe by
density operators. A pure state is a unit vector $|\psi\rangle$
in $\mathcal{H}$, which corresponds to a density operator $\rho= |\psi
\rangle \langle\psi|$, the projection operator on~$|\psi\rangle$.
An ensemble of pure states corresponds to the case that the quantum
system is in one of
states $|\psi_k \rangle$, $k=1,\ldots, K$, with probability $p_k$
being in state
$|\psi_k \rangle$, and the corresponding density operator is
%
%e1 ###
\begin{equation} \label{rho}
\rho= \sum_{k=1}^K p_k |\psi_k \rangle \langle\psi_k |.
\end{equation}
Let $|\psi(t)\rangle$ be the state of the quantum system at time $t$.
The states $|\psi(t_1)\rangle$
and $|\psi(t_2)\rangle$ at $t_1$ and $t_2$ are connected through
$ |\psi(t_2)\rangle= U(t_1, t_2) |\psi(t_1)\rangle, $ where
$U(t_1,t_2)$ is a unitary operator
depending only on time $t_1$ and $t_2$.
In fact, there exists a Hermitian operator $H$, which is known as the
Hamiltonian of the
quantum system, such that
$ U(t_1,t_2) = \exp[ -i H (t_2-t_1)]. $
With Hamiltonian $H$, we may depict
the continuous time evolution of $|\psi(t)\rangle$ by Schr\"
odinger's equation
%
%e2 ###
\begin{equation} \label{schrodinger1}
i \frac{\partial|\psi(t) \rangle}{\partial t} = H |\psi(t) \rangle.
\end{equation}
See Holevo (\citeyear{H1982}) and Sakurai (\citeyear{S1995}).

%s2.2 ###
\subsection{Quantum probability}\label{sec2.2}
Quantum mechanics can be tested by checking its predications with
experiments of
performing measurements on quantum systems.
The common quantum measurements are on observables such as
position, momentum, spin and so on, where an \textit{observable} is defined
as a Hermitian operator on Hilbert space $\mathcal{H}$. Consider an
observable $\bX$
with a~discrete spectrum so that it can be written in a diagonal form
%
%e3 ###
\begin{equation} \label{diagonal}
\bX= \sum_{a=1}^p x_a \bQ_a,
\end{equation}
where $x_a \in \mathbb{R}$ are eigenvalues of $\bX$ and $\bQ_a$ are the corresponding
one-dimensional projections onto the eigenvectors of $\bX$.
Possible measurement outcomes of the observable are described by
measure space
$(\Omega, \mathcal{F})$. For a~quantum system with
state $\rho$, the result of the measurement is random with probability
distribution $P_\rho$ over $(\Omega, \mathcal{F})$. We denote by $X$ the
result of
the measurement of observable $\bX$ given by (\ref{diagonal}).
The result $X$ is a random variable
and takes values in $\Omega= \{x_1, x_2, \ldots\}$. With a quantum system
prepared in the state $\rho$, the result $X$ has a probability distribution
$ P_\rho[ X = x_a] = \operatorname{tr}( \rho \bQ_a). $
%with expectation $E_\rho[X] = \sum_{a=1}^p x_a P_\rho[ X = x_a] = \operatorname{tr}(
With the probability distribution $P_\rho$, we can easily derive the
expectation and variance
of $\bX$ in the state
\begin{eqnarray*}
E_\rho[\bX] &=& \operatorname{tr}( \rho \bX) = \sum_{a=1}^p x_a P_\rho[ X = x_a]
= E_{P_\rho} (X),\\
\operatorname{Var}_\rho[\bX] &=& \operatorname{tr}[ \rho\bX^2] - [\operatorname{tr}( \rho \bX)]^2.
\end{eqnarray*}
Measuring the outcomes of observable $\bX$ will alter the state of the
quantum system.
If the state of the quantum system is $\rho$ immediately before the
measurement, then
the probability that the result $x_a$ occurs is $P_\rho[ X = x_a] =
\operatorname{tr}( \rho \bQ_a)$
and the state of the system after the measurement result $x_a$ is equal to
$ \bQ_a \rho \bQ_a /\operatorname{tr}(\bQ_a \rho \bQ_a). $
Similarly, using the spectral theory of self-adjoint operators, we may
describe observables
with continuous spectrum and continuous measurement outcomes.
See Barndorff-Nielsen, Gill and Jupp (\citeyear{BGP2003}) and Holevo (\citeyear{H1982}).

%s2.3 ###
\subsection{Quantum computation}\label{sec2.3}
Quantum systems can be simulated via computers, but quantum simulation
requires enormous computational resources.
Classic computers may have great difficulty to efficiently simulate
general quantum systems,
while quantum computers built upon quantum systems are ideal for
quantum simulation.

Analog to the fundamental concept of the bit in classical computation
and classical information,
we have quantum bit in quantum computation and quantum information and
call it qubit
for short. Just like a classical bit with state either $0$ or $1$, a
qubit has states $|0\rangle$
and $|1\rangle$.
However, there is a real difference between a bit and a qubit.
Besides states $|0\rangle$ and $|1\rangle$, a qubit can also take
states as their superpositions,
which are the linear combinations of $|0\rangle$ and $|1\rangle$,
\[
|\psi\rangle= \alpha_0 |0\rangle+ \alpha_1 |1\rangle,
\]
where complex numbers $\alpha_0$ and $\alpha_1$ are called amplitudes
satisfying
$|\alpha_0|^2 +|\alpha_1|^2=1$.
In other words, the states of a qubit are unit vectors in a
two-dimensional complex vector space,
and states $|0\rangle$ and $|1\rangle$ consist of an orthonormal
basis for the space and are
often referred to as computational basis states. The qubit is the
simplest quantum system.
Unlike a classical bit which can be examined
to determine whether it is in the state $0$ or $1$, for a qubit we can
not determine its state
and find the values of $\alpha_0$ and $\alpha_1$ by examining it.
Quantum mechanics shows that
we can measure a qubit and obtain either the result $0$, with
probability $|\alpha_0|^2$, or
the result $1$, with probability $|\alpha_1|^2$. A qubit can be
actually realized as physical objects
in many different physical systems, such as the two
different polarizations of a~photon, the alignment of a nuclear spin in
a uniform magnetic field
or two states of an electron orbiting a single atom. In the atom model
case, we may correspond
$|0\rangle$ and $|1\rangle$ with the so-called ``ground'' or
``excited''
states of the electron,
respectively. As the atom is shined by light with suitable energy and
for a proper amount of time,
we can move the electron from the $|0\rangle$ state to the $|1\rangle
$ state and vice versa. Moreover,
by shortening the length of time shining the light on the atom, we can
move an electron initially
in the state~$|0\rangle$ to ``halfway'' between $|0\rangle$ and
$|1\rangle$, say, into a state
$|+ \rangle=(|0\rangle+ |1\rangle)/\sqrt{2}$.

Like classical bits, we may consider multiple qubits. The states of two
qubits are unit vectors
in a four-dimensional complex vector space, with four computational
basis states labeled by
$|00\rangle$, $|01\rangle$, $|10\rangle$ and $|11\rangle$.
In general, a system of $b$ qubits has $2^b$ computational basis states
of the form $|x_1 x_2\cdots x_b \rangle$,
$x_j=0$ or $1$, $j=1,\ldots, b$, that generate a $2^b$-dimensional
complex vector space, and a superposition
state in the system is specified by $2^b$ amplitudes. As $2^b$
increases exponentially in $b$, it is easy for such a
system to have an enormously big vector space.
A quantum system consisting of even a few dozens of ``qubits'' will
strain the resources of even the largest supercomputers.
Consider a system with $50$ qubits. $2^{50} \approx10^{15}$ complex
amplitudes are needed to depict its quantum state.
With $128$ bits of precision, it requires approximately $32$ thousand
terabytes of information to store all $10^{15}$
complex amplitudes. Had Moore's law continued on schedule, such storage
capacity would be available in supercomputers
during the second decade of the twenty-first century.
A system with $b=500$ qubits has the number of amplitudes
larger than the estimated number of atoms in the universal. It is
unimaginable to store all $2^{500}$ complex numbers in any classical
computers. In principle, a quantum system
with only a few hundred atoms can manage such an enormous amount of
data and execute calculations as the system
evolves. Quantum computation and quantum information are to find ways
to utilize the immense potential
computational power in quantum systems. See Clarke and Wilhelm (\citeyear{CW2008}),
Deutsch (\citeyear{D1985}), DiCarlo et al. (\citeyear{Detal2009}),
Feynman (\citeyear{F1982}), Lloyd (\citeyear{L1996}), DiVincenzo (\citeyear{D1995}), Nielsen and Chuang (\citeyear{NC2000}) and
Shor (\citeyear{S1994}).

%s3 ###
\section{Statistical analysis of quantum simulation}\label{sec3}

The key for the simulation of a quantum system lies in the solution of
Schr\"odinger's
equation (\ref{schrodinger1}) which governs the dynamic evolution of
the system, and
the quantum simulation can be done via either classic computing or
quantum computing.
Schr\"odinger's equation for a typical Hamiltonian with real particles
usually consists
of elliptical differential equations, where each differential equation
can be easily simulated by a classical computer.
The real challenge in stimulating a quantum system is to solve the
exponential number of such
differential equations. Consider a quantum system that is described by
$b$ qubits.
As $b$ qubits have $2^b$ amplitudes, for stimulating the dynamic
behavior of $b$ qubits evolving according to Schr\"odinger's
equation, a system of
$2^b$ differential equations must be solved. Because of the exponential
growth in the number
of differential equations, simulating quantum systems by classical
computers is feasible only for
special cases where insightful approximations are available to
dramatically reduce the
effective number of differential equations involved.
Quantum computers may be ideal for the simulation of naturally
occurring quantum systems.
See Abrams and Lloyd (\citeyear{AL1997}), Boghosian and Taylor (\citeyear{BT1998}), Feynman (\citeyear{F1982}),
Lloyd (\citeyear{L1996}) and Zalka (\citeyear{Z1998}).
Whether quantum simulation is via classic computing or quantum
computing, its
statistical aspect essentially remains the same.

%s3.1 ###
\subsection{Simulate a quantum system}\label{sec3.1}

The heart of quantum simulation is to solve Schr\"odinger equation
(\ref{schrodinger1})
which has the solution
%
%e4 ###
\begin{equation} \label{schrodinger2}
| \psi(t) \rangle= e^{-i H t} | \psi(t_0) \rangle.
\end{equation}
Numerical evaluation of $e^{-i H t}$ is needed.
The Hamiltonian $H$ is usually exponentially large and extremely
difficult to
exponentiate. The common approach in numerical analysis that uses the
first-order
linear approximation, $1 - i H \delta$, of $e^{-i H (t + \delta)} -
e^{-i H t}$ often yields unsatisfactory numerical solutions.

Efficient evaluation of the solutions (\ref{schrodinger2}) with high
order approximation
exists for many classes of Hamiltonians. For most physical systems the
Hamiltonians involve
only location interactions, which originate from the fact that most
interactions fall off
with increasing distance or difference in energy. Specifically, a
system of $\alpha$
particles in a $d$-dimensional space often has a~Ha\-miltonian of the form
%
%e5 ###
\begin{equation} \label{Hamiltonian}
H = \sum_{\ell=1}^L H_\ell,
\end{equation}
where $L$ is a polynomial in $\alpha+d$, and each $H_\ell$ acts on a
small subsystem of finite
size free from $\alpha$ and $d$. Typical examples of the terms $H_\ell
$ are one-body Hamiltonians
and two-body interactions
such as the ones in the Hubbard and Ising models [Altland and Simons
(\citeyear{AS2006}) and Dziarmaga (\citeyear{D2005})]. As a~result,
$e^{-i H_\ell\delta}$ is easy to compute numerically,
although $e^{-i H \delta}$ is very hard to evaluate. Because $H_\ell$
and $H_k$ are
noncommutable,
$e^{-i H \delta} \neq e^{-i H_1 \delta} \cdots e^{-i H_L \delta}$. Using
the Trotter formula [Kato (\citeyear{K1978}), Sornborger and Stewart (\citeyear{SS1999}) and
Trotter (\citeyear{T1959})], we approximate $e^{-i H \delta}$ by $U_{\delta}$ which
requires only the evaluation of each $e^{-i H_\ell\delta}$,
where
%
%e6 ###
\begin{equation} \label{eqU}
U_{\delta} = [ e^{-i H_1 \delta/2 } \cdots e^{-i H_L \delta/2 } ]
[ e^{-i H_L \delta/2 } \cdots e^{-i H_1 \delta/2 } ].
% U_{\delta} = [ e^{-i H_1 \delta} \cdots e^{-i H_L \delta} ]
% [ e^{-i H_L \delta} \cdots e^{-i H_1 \delta} ].
\end{equation}
%
%H_L \Delta t}
% \cdots e^{-i H_1 \Delta t}. \]
%Show that the restriction of $H_k$ to involve at most $c$ particles
%implies that in the
%sum of (4.97), $L$ is upper bounded by a polynomial in n. $n^c$
%The heart of quantum simulation algorithms is the following asymptotic
%approximation theorem

Suppose that the quantum system starts at $t_0$ with initial state
$|\psi(t_0)\rangle$ and ends at final time $T$. For an integer $m$,
let $\delta= T/m$ and $t_j=j \delta$, $j=0,1,\ldots, m$. The quantum
simulation is to apply approximation $U_\delta$ of $e^{-i H \delta}$ to
solutions (\ref{schrodinger2}) at $t_j$ iteratively and generate approximate
solutions for $|\psi(t_j)\rangle$.
Denote by $|\tilde{\psi}(t_j) \rangle$ the state at $t_j$ obtained
from the quantum simulation as
an approximation of the true state $|\psi(t_j)\rangle$ at $t_j$. Then
for $j=1,\ldots, m$,
\begin{eqnarray} \label{finalstate}
|\psi(t_{j})\rangle&=& e^{-i H \delta} |\psi(t_{j-1})\rangle= e^{-i
H j \delta} |\psi(t_0)\rangle,\nonumber\\[-8pt]\\[-8pt]
|\tilde{\psi}(t_{j})\rangle &=& U_\delta |\tilde{\psi
}(t_{j-1})\rangle= U^j_\delta |\psi(t_0) \rangle.\nonumber
% j = 1,\ldots, m.
\end{eqnarray}

If the initial state of the quantum simulation is a pure state $|\psi
(t_0)\rangle$,
then the true final state and the simulated final state are also pure states
$|\psi(t_{m})\rangle$ and $|\tilde{\psi}(t_m)\rangle$,
respectively, with
corresponding density operators
\[
\rho(t_0) = |\psi(t_m) \rangle \langle\psi(t_m)|,\qquad
\tilde{\rho} = |\tilde{\psi}(t_m) \rangle \langle\tilde{\psi
}(t_m)|.
\]
When the initial state is an ensemble of pure states, with
probability $p_k$ being pure state $|\psi_k(t_0)\rangle$, $k=1,\ldots
, K$, and
corresponding density operator
\[
\rho(t_0) = \sum_{k=1}^K p_k |\psi_k(t_0) \rangle \langle\psi_k(t_0)|,
\]
then at time $t_j$ the true state and the simulated state are also
ensembles of
pure states with respective density operators
\begin{eqnarray*}
\rho(t_j) &=& \sum_{k=1}^K p_k |\psi_k(t_j) \rangle \langle\psi
_k(t_j)|,
\\
\tilde{\rho}(t_j) &=& \sum_{k=1}^K p_k |\tilde{\psi}_k(t_j) \rangle
\langle
\tilde{\psi}_k(t_j)|,
\end{eqnarray*}
where for $k=1,\ldots, K$,
\begin{eqnarray*}
|\psi_k(t_j) \rangle&=& e^{-i H \delta} |\psi_k(t_{j-1})\rangle=
e^{-i H j \delta} |\psi_k(t_0) \rangle,\\
|\tilde{\psi}_k(t_j) \rangle&=& U_\delta |\tilde{\psi
}_k(t_{j-1})\rangle=
U_\delta^j |\psi_k(t_0) \rangle.
\end{eqnarray*}

%s3.2 ###
\subsection{Monte Carlo quantum simulation}

Quantum simulation provides an excellent way for the study of
complex phenomena in physical and biology systems and evaluating
hard-to-obtain quantities
in the system. Examples include the dielectric constant, the mass of
the proton, conductivity,
magnetic susceptibility of materials and molecules in biological
systems. As the results of
quantum measurement outcome are random, repeated measurements need to
be performed in order
to obtain reliable estimators of the quantities. In the quantum setup,
a quantity of interest
is of the form
%
%e7 ###
\begin{equation} \label{theta}
\theta= E_\rho(\bX) = \operatorname{Tr}(\bX \rho) = E_{P_\rho} (X),
\end{equation}
where $\bX$ is an observable, $X$ is its measurement result, and $\rho
$ is the state of the quantum
system under which we perform the measurements and evaluate the
quantity $\theta$.

A Monte Carlo quantum simulation method is designed to estimate $\theta
$ as follows.
We prepare the quantum system at initial state $\rho(t_0)$ and make it evolve
to final state $\rho(t_m)=\rho$. The quantum simulation procedure
described in Section \ref{sec3.1} is used to simulate
the evolutions of the quantum system from initial state
$\rho(t_0)$ to final state $\rho(t_m)$ according to Schr\"odinger's equation
(\ref{schrodinger2}) with some Hamiltonian $H$ of the form given by
(\ref{Hamiltonian}). We
repeatedly perform the measurements of observable $\bX$ in such $n$
identically simulated
quantum systems at the simulated final state and obtain measurement
results $X_1,\ldots, X_n$. We estimate $\theta$ defined in (\ref
{theta}) by
%
%e8 ###
\begin{equation} \label{thetahat}
\hat{\theta} = \frac{1}{n} \sum_{j=1}^n X_j.
\end{equation}

Since measurements and state approximations in quantum simulation are
involved with random
fluctuations and systematic errors, $\hat{\theta}$ as a Monte Carlo
estimator of
$\theta$ has variance and bias. We use mean square error (MSE) criterion
to gauge its performance. The Monte Carlo quantum simulation method for
obtaining
estimator $\hat{\theta}$ requires to repeat the
simulation of the quantum system $n$ times, and each simulation needs
to calculate $m$
approximations of states at $t_j$, $j=1,\ldots, m$.
The whole Monte Carlo quantum simulation procedure needs to make total
$N=m n$ state
approximations for the quantum system. One important problem is to
determine the strategy to allocate $m$ and $n$ with given $N = m n$
so that the MSE of $\hat\theta$ is minimized. We derive the MSE of
$\hat{\theta}$ and
study the problem in the following theorem.

\begin{thm} \label{thm1}
For a quantum system evolving in time interval $[0, T]$, assume that
its Hamiltonian $H$
and observable $\bX$ satisfy (\ref{Hamiltonian}) in Section \ref{sec3.1} and
(\ref{condition1}) in Section \ref{sec3.3}, respectively. Then
\[
E [(\hat{\theta} - \theta)^2 ] \leq\frac{C_1}{n} + \frac
{C_2}{m^4} =
\frac{C_1}{N \delta} + \frac{C_2 \delta^4}{T^4},
\]
where $C_1$ and $C_2$ are generic constants free from $m$ and $n$.
Thus, when $n = C_1 m^4/C_2$ and $m = (C_2/C_1)^{1/5} N^{1/5}$, the MSE
bound is
asymptotically minimized and
\[
E [(\hat{\theta} - \theta)^2 ] \leq C_1^{4/5} C_2^{1/5} N^{-4/5}.
\]
\end{thm}

The theorem indicates that the MSE of $\hat{\theta}$ is of order $C_1
n^{-1} + C_2 m^{-4}$,
where from the proof of the theorem in Section \ref{sec3.3} below we see that
$C_1$ is the variance of $\bX$ and
$C_2$ is the difference of the expectations of $\bX$ under true state
$\rho(t_m)$ and the
simulated state $\tilde{\rho}(t_m)$. As the Monte Carlo quantum
simulation procedure performs
$n$ repeated simulations of the quantum system with $m$ state
approximations for each simulation,
if we have the computational capacity of carrying out a total of
$N=m n$ state approximations in the Monte Carlo quantum simulation, the
theorem provides
optimal strategy for the allocation of $m$ and $n$ that
%$ n = C_1 m^4/C_2, m = (C_2/C_1)^{1/5} N^{1/5}, $
minimizes the MSE of~$\hat{\theta}$.

%s3.3 ###
\subsection{\texorpdfstring{Proof of Theorem \protect\ref{thm1}}{Proof of Theorem 3.1}}\label{sec3.3}

As usual, the MSE analysis involves deriving its variance and bias,
%
%e9 ###
\begin{equation} \label{eq1}
E[(\hat{\theta} - \theta)^2]= \operatorname{Var}(\hat{\theta}) + (E \hat{\theta
} - \theta)^2.
\end{equation}
We need to fix some notation to facilitate further analysis. The target
$\theta$ is
defined under the true final state $\rho(t_m)$, while the quantum
simulation is under
approximate final state
$\tilde{\rho}(t_m)$ which is close to $\rho(t_m)$. To make the
problem realistic,
we impose the following assumption to ensure that observable $\bX$
behaves well in states
close to the true final state. With the true final state of the form
\[
\rho(t_m) = \sum_{k=1}^K p_k |\psi_k(t_m) \rangle\langle\psi
_k(t_m) |,
\]
we assume that for some small $\eta>0$
%
%e10 ###
\begin{equation} \label{condition1}
\max_{1 \leq k \leq K}
\sup\bigl\{ \|\bX|\phi\rangle\|, \bigl\| |\phi\rangle- | \psi
_k(t_m)\rangle\bigr\| < \eta\bigr\} < \infty.
\end{equation}
The condition is to ensure that
observable $\bX$ has two finite moments under states in a small
neighborhood of the
true state $\rho(t_m)$ of the quantum system.

Since a simple conditional argument will reduce the proof from the
general ensemble state
case to the pure state
case, for simplicity, we consider the Monte Carlo study with pure
states. The state used
in (\ref{theta}) is the pure state $|\psi(t_m) \rangle$ or $\rho=
|\psi(t_m)\rangle
\langle\psi(t_m)|$, under which observable $\bX$ and its measurement
result $X$ are
considered.
The measurement results $X_1,\ldots, X_n$ are obtained from the
quantum simulation
under the simulated state $|\tilde{\psi}(t_m) \rangle$ or $\tilde
{\rho}= |\tilde{\psi}(t_m)\rangle
\langle\tilde{\psi}(t_m)|$. Therefore, to analyze the bias and
variance, we need to evaluate the
expectations and variances of $X_i$
under $\tilde{\rho}$ but compute the corresponding quantities of $X$
under $\rho$.
%&& E[(\hat{\theta} - \theta)^2]= \operatorname{Var}(\hat{\theta}) + (E \hat{\theta} -
%&& = \frac{1}{n} \operatorname{Var}(X_1) + \{\operatorname{Tr}(\bX\tilde{\rho}) - \operatorname{Tr}(\bX\rho)\}^2.
%&& = \frac{1}{n} \operatorname{Var}(X_1) + \{\operatorname{Tr}[\bX(\tilde{\rho} -\rho)]\}^2 \\

The bias $E \hat{\theta} - \theta= \operatorname{Tr}(\bX\tilde{\rho}) - \operatorname{Tr}(\bX
\rho)$
is due to the differences between~$|\tilde{\psi}(t_j) \rangle$
obtained in the quantum simulation and the true quantum states~$|\psi
(t_j)\rangle$.
We will prove in Proposition \ref{prop2} below
%
%e11 ###
\begin{equation} \label{eq2}
\bigl\| | \tilde{\psi}(t_m) \rangle- | \psi(t_m) \rangle\bigr\| \leq C
\delta^2.
\end{equation}
Thus, we derive the bias
%
%e12 ###
\begin{eqnarray} \label{eq3}
| E \hat{\theta} - \theta| &=& | \operatorname{Tr}(\bX\tilde{\rho}) - \operatorname{Tr}(\bX
\rho) | \nonumber\\
& =& \bigl| \langle\tilde{\psi}(t_m)| \bX| \tilde{\psi}(t_m) \rangle
- \langle\psi(t_m)| \bX| \psi(t_m) \rangle\bigr| \nonumber\\
& \leq&\bigl| \langle\tilde{\psi}(t_m) - \psi(t_m) | \bX| \tilde{\psi
}(t_m) \rangle\bigr| +
\bigl| \langle\tilde{\psi}(t_m)| \bX| \tilde{\psi}(t_m) - \psi(t_m)
\rangle\bigr| \\
&\leq&\bigl\| | \tilde{\psi}(t_m) - \psi(t_m) \rangle\bigr\| \bigl(\| \bX|
\tilde{\psi}(t_m) \rangle\|
+ \| \bX|\psi(t_m) \rangle\| \bigr) \nonumber\\
& \leq& C \delta^2,\nonumber
\end{eqnarray}
where the last inequality is due to (\ref{eq2}) and condition (\ref{condition1}).
The variance of~$\hat{\theta}$ is easy to obtain
%
%e13 ###
\begin{equation} \label{eq4}
\operatorname{Var}(\theta) = \frac{1}{n} \operatorname{Var}(X_1) = \frac{1}{n} \{\operatorname{Tr}( \bX^2 \tilde
{\rho}) - [\operatorname{Tr}(\bX\tilde{\rho})]^2
\}.
\end{equation}
As we have shown above, the $\operatorname{Tr}( \bX \tilde{\rho})$ approach to
$\theta=\operatorname{Tr}( \bX \rho)$
as $m \rightarrow\infty$ or, equivalently, $\delta\rightarrow0$. As for
$\operatorname{Tr}( \bX^2 \tilde{\rho})$,
\[
|\operatorname{Tr}( \bX^2 \tilde{\rho})|= | \langle\tilde{\psi}(t_m) |\bX^2
|\tilde{\psi}(t_m)\rangle|
= \| \bX|\tilde{\psi}(t_m)\rangle\|^2,
\]
whose finiteness is a consequence of (\ref{eq2}) and condition (\ref
{condition1}).
Collecting together (\ref{eq1}), (\ref{eq3}) and (\ref{eq4}), we conclude
\[
E[(\hat{\theta} - \theta)^2] \leq\frac{C_1}{n} + C_2 \delta^4
\sim\frac{C_1}{n} +
\frac{C_2 T^4}{m^4},
\]
which is asymptotically minimized when $n \sim m^4 \sim N^{4/5}$. To
complete the proof of
Theorem \ref{thm1}, we show (\ref{eq2}) in the rest of the section.

The quantum simulation uses $|\tilde{\psi}(t_j) \rangle$ to
approximate the true quantum states
$|\psi(t_j)\rangle$ and thus results in approximation errors. We
define the following quantity to
measure the approximation errors. Suppose $U_1$ and $U_2$ are two
unitary transformations, the
operator norm of the difference between~$U_1$ and $U_2$,
\[
\Gamma(U_1, U_2) = \max_{\|\phi\|=1} \bigl\| (U_1 - U_2) |\phi\rangle\bigr\|,
\]
is used to measure the closeness of $U_1$ and $U_2$.

In the quantum simulation scheme, we approximate $e^{-i H j \delta}$
by $U_\delta^j$
defined in (\ref{eqU}). Naturally we use $\Gamma(U^j_\delta, e^{-i H
j \delta})$ to
gauge the approximation errors in the quantum simulation. Below we
derive the
order in terms of $\delta$ for approximation errors $U_{\delta}^j -
e^{-i H j \delta}$ in
the quantum simulation.
\begin{prop} \label{prop2}
The following inequality holds uniformly for $j =\break 1, \ldots, m$:
\[
\Gamma( U^j_{\delta}, e^{-i H j \delta} ) \leq C L \delta^2.
\]
\end{prop}
\begin{pf}
First we prove the inequality for $j=1$.
For the case of \mbox{$L=2$}, $H=H_1+H_2$. Expanding exponential functions of
$H_i$ and
using simple algebraic manipulation, we have
\begin{eqnarray*}
%&& e^{-i (H_1+H_2) \delta} = e^{-i H_1 \delta/2} e^{-i H_2 \delta}
%e^{-i H_1 \delta/2} + O(\delta^3) \\
e^{-i H_1 \delta/2} &=& I - \frac{i}{2} H_1 \delta- \frac{1}{8}
H_1^2 \delta^2 + O(\delta^3), \\
e^{-i H_2 \delta} &=& I - i H_2 \delta- \frac{1}{2} H_2^2 \delta^2
+ O(\delta^3),\\
e^{-i (H_1+H_2) \delta}&=&I - i (H_1+H_2) \delta- \frac{1}{2}
(H_1+H_2)^2 \delta^2 + O(\delta^3)\\
&=&I - i (H_1 + H_2) \delta- \frac{1}{2} \delta^2 (H_1 H_2 + H_2 H_1
+ H^2_1 + H^2_2) + O(\delta^3) \\
&=& \biggl( I - \frac{i}{2} H_1 \delta- \frac{1}{8} H_1^2 \delta^2 \biggr) \biggl(I
- i H_2 \delta
- \frac{1}{2} H_2^2 \delta^2 \biggr)\\
&&{}\times \biggl(I - \frac{i}{2} H_1 \delta- \frac
{1}{8} H_1^2
\delta^2 \biggr) + O(\delta^3) \\
&=& e^{-i H_1 \delta/2} e^{-i H_2 \delta} e^{-i H_1 \delta/2} +
O(\delta^3) = U_\delta+
O(\delta^3).
\end{eqnarray*}
For general $L$, let $H_j^* = \sum_{\ell=j}^L H_\ell$. Then $H =
H^*_1$ and $H_j^*=H_j + H^*_{j+1}$ for
$j=1,\ldots, L-1$. We repeatedly apply the above result for the case
of $L=2$ to the case of
$H_j^*=H_j + H^*_{j+1}$ and obtain
\begin{eqnarray*}
%&& e^{-i H \delta} = e^{-i (H_1+H_2 + \cdots+ H_L) \delta} = e^{-i
%(H_1+ H^*_2) \delta} \\
e^{-i H \delta} &=& e^{-i (H_1+ H^*_2) \delta} \\
&=& e^{-i H_1 \delta/2} e^{-i H^*_2 \delta} e^{-i H_1 \delta/2} +
O(\delta^3) \\
&=& e^{-i H_1 \delta/2} e^{-i (H_2 + H^*_3) \delta} e^{-i H_1 \delta
/2} + O(\delta^3) \\
&=& e^{-i H_1 \delta/2} [ e^{-i H_2 \delta/2} e^{-i H^*_3 \delta}
e^{-i H_2 \delta/2}
+ O(\delta^3) ] e^{-i H_1 \delta/2} + O(\delta^3) \\
&=& e^{-i H_1 \delta/2} e^{-i H_2 \delta/2} e^{-i H^*_3 \delta}
e^{-i H_2 \delta/2}
e^{-i H_1 \delta/2} + O(2 \delta^3) \\
&=& \cdots= U_\delta+ O(L \delta^3), % \diamondsuit
\end{eqnarray*}
which implies the inequality for $j=1$.

Second we show the inequality for $j=2$.
Let $V_\delta=e^{-i H \delta}$.
%Proposition \ref{prop1} indicates that the inequality holds for $j=1$.
For any state~$|\phi\rangle$,
\begin{eqnarray*}
\bigl\|(U^2_\delta- V^2_\delta)|\phi\rangle\bigr\| &\leq&\bigl\|(U^2_\delta-
U_\delta V_\delta)|\phi\rangle\bigr\|
+ \bigl\|(U_\delta V_\delta- V^2_\delta)|\phi\rangle\bigr\| \\
& \leq&\bigl\|U_\delta (U_\delta- V_\delta)|\phi\rangle\bigr\|
+ \bigl\|(U_\delta- V_\delta) V_\delta|\phi\rangle\bigr\| \\
& \leq&\bigl\|(U_\delta- V_\delta)|\phi\rangle\bigr\|
+ \bigl\|(U_\delta- V_\delta)|\phi^\prime\rangle\bigr\|,
\end{eqnarray*}
where $|\phi^\prime\rangle=V_\delta|\phi\rangle$.
Since $V_\delta$ is unitary, $\| |\phi^\prime\rangle\|= \||\phi
\rangle\|$. On both sides of the above
inequality we take the maximum over all $\phi$ with $\| |\phi\rangle
\|=1$ and obtain
\begin{eqnarray*}
\Gamma(U_\delta^2, V_\delta^2) &=& \max_{\| |\phi\rangle\|=1} \bigl\|
(U^2_\delta- V^2_\delta)|\phi\rangle\bigr\| \\
& \leq&2 \max_\phi\bigl\|(U_\delta- V_\delta)|\phi\rangle\bigr\| \\
& = &2 \Gamma(U_\delta, V_\delta) \leq C L \delta^3,
\end{eqnarray*}
where the last inequality is from the proved case of $j=1$.
%Proposition \ref{prop1}.

The result for general $j$ follows by induction. Since $U_\delta$ and
$V_\delta$ are unitary, we repeatedly
apply the same technique for proving the case of $j=2$ as follows. For
$|\phi\rangle$ with
$\| |\phi\rangle\|=1$,
\begin{eqnarray*}
\bigl\|(U^j_\delta- V^j_\delta)|\phi\rangle\bigr\| &\leq&\bigl\|(U^j_\delta-
U^{j-1}_\delta V_\delta)|\phi\rangle\bigr\|
+ \bigl\|(U^{j-1}_\delta V_\delta- V^j_\delta)|\phi\rangle\bigr\| \\
& \leq&\bigl\|U^{j-1}_\delta (U_\delta- V_\delta)|\phi\rangle\bigr\|
+ \bigl\|(U^{j-1}_\delta- V^{j-1}_\delta) V_\delta|\phi\rangle\bigr\| \\
& \leq&\bigl\|(U_\delta- V_\delta|\phi\rangle\bigr\|
+ \bigl\|(U^{j-1}_\delta- V^{j-1}_\delta) |\phi^\prime\rangle\bigr\|,
\end{eqnarray*}
where $|\phi^\prime\rangle=V_\delta|\phi\rangle$. Taking the
maximum over all $|\phi\rangle$ with
$\| |\phi\rangle\|=1$ on both sides of the above inequality, we get
\begin{eqnarray*}
\Gamma(U_\delta^j, V_\delta^j)& \leq&\Gamma(U_\delta, V_\delta)
+ \Gamma(U^{j-1}_\delta, V^{j-1}_\delta)
\leq\cdots\leq j \Gamma(U_\delta, V_\delta) \\
& \leq& m \Gamma(U_\delta, V_\delta) \leq C L \delta^2,
\end{eqnarray*}
where the last inequality is from $m \delta=T$ and the proved case of $j=2$.
%Proposition \ref{prop1} $\diamondsuit$
\end{pf}

%Specifically, for a system of $n$ particles,
%where each $H_j$ acts on at most constant a constant $c$ number of
%systems, and $L$ is a polynomial in $N$'.
%(N = b qubits).

%s4 ###
\section{An example}\label{sec4}

There are a few interesting and realistic quantum systems such as the
quantum Ising model and simple
harmonic oscillator for which some analytic solutions are available
[Dziarmaga (\citeyear{D2005}) and Sakurai
(\citeyear{S1995})]. In this section we illustrate the Monte Carlo quantum
simulation method with simple harmonic
oscillator.

%s4.1 ###
\subsection{Three-dimensional isotropic harmonic oscillator}
$\!\!\!$We consider a quantum system of $d/3$ particles with three-dimensional isotropic
harmonic oscillator. This is a $d$-dimensional quantum system. By using
the natural scales of
length and energy in terms of particle mass, angular frequency and
Planck's constant,
we have the following simple expression for the Hamiltonian of the system:
\[
H = (\bXi^2 - \bDelta)/2,
\]
where harmonic operator $\Delta$ and isotropic multiplication operator
$\bXi^2$ are defined as follows:
\begin{eqnarray*}
\bDelta&=& \sum_{j=1}^d \nabla^2_{j},\qquad  \bXi^2 = \sum_{j=1}^d \bolds{\xi}_j^2,\qquad\nabla_{j}=\frac{\partial}{\partial x_{j}},
\\
{}[\bolds{\xi}_{j} f](\mathbf{x}) &=& x_{j} f(\mathbf{x}),\qquad
\mathbf{x}= (x_1,\ldots, x_d)^\dagger\in \mathbb{R}^{d}, f \in L(\mathbb{R}^{d}),
\end{eqnarray*}
and for $\ell=1,\ldots, d/3$, $(x_{3 \ell-2}, x_{3 \ell-1}, x_{3
\ell})$ specify
the position coordinates of the $\ell$th particle in $\mathbb{R}^3$.
%The canonical commutation relations between the operators $\nabla_j$
%and $\bX_j$ are
% [\bX_{j}, \bX_{k}]=[\nabla_{j},\nabla_{k}]=0 \]
%x_j^2} ) = \frac{1}{2}
% (\bR^2 - \Delta) \]
%Operator $\bR$ is the position operator by multiplying Euclidean
%distance.
Hamiltonian $H$ can be written as a sum of $d$ local Hamiltonians
% = ( \bxi_j^2 - \nabla_j^2 )/2, \]
%with one dimensional harmonic oscillator,
%where
$H_j= ( \bolds{\xi}_j^2 - \nabla_j^2 )/2$. $H_j$ are
one-dimensional harmonic oscillators and have expression
$H_j = A^+ A^- + 1/2$, where
$A^+$ and $A^-$ are creation and
annihilation operators given below,
\[
A^{\pm}= (\bolds{\xi}_j \mp\nabla_j)/\sqrt{2},\qquad  %[
[A^-, A^+] = A^- A^+ - A^+ A^- = I.
\]
As shown in Dziarmaga (\citeyear{D2005}), the Hamiltonian of the quantum Ising
model can also be
expressed by similar product of creation and annihilation operators.

Operator $A^+ A^-$ has eigenvalues $k$ for $k=0, 1, \ldots,$ with
eigenfunctions
defined by normalized Hermite polynomials,
%
%e14 ###
\begin{equation} \label{hermite1}
h_k(x) = \frac{(-1)^k}{\sqrt{2^k k!\sqrt{\pi}}} e^{x^2/2} \frac
{d^k}{dx^k} ( e^{-x^2} ).
% \psi_k(x)= \frac{1}{\sqrt{2^n n!\sqrt{\pi}}} h_k(x) \exp(-x^2/2 ), k
%= 0, 1,\ldots,
\end{equation}
%
%where $h_k(x)$ are Hermite polynomials
In fact, it can be directly verified that %$A^- A^- \h_k(x)= k h_k(x)$,
$[x^2 h_k(x) - h^{\prime\prime}_k(x)]/2 = (k+1/2) h_k(x)$
and, thus,
$h_k(x)$ are eigenfunctions of $H_j$ corresponding to eigenvalues $k+1/2$.
As Hamiltonian $H$ is a sum of $d$ one-dimensional harmonic
oscillators, eigenfunctions of Hamiltonian $H$ are
given by
\[
h_{\vec{k}}(\mathbf{x})= \prod_{j=1}^d h_{k_j}(x_j),\qquad  \vec
{k}=(k_1,\ldots, k_d)^\dagger,
k_j=0, 1,\ldots,
\]
with corresponding eigenvalues $\sum_{j=1}^d k_j + d/2$.

%s4.2 ###
\subsection{Quantum simulation}
To make the quantum simulation manageable computationally, we consider
the simulation of the following six-dimensional
quantum system described by $12$ qubits. It requires a Hilbert space of
dimension $2^{12}=4096$ to accommodate the quantum
system. We use the first four eigenfunctions of a one-dimensional
harmonic oscillator to form two qubit states in
each dimension. With the six sets of the four eigenfunctions, we obtain
$4096$ eigenfunctions of product form
and generate a Hilbert space of dimension $4096$ to accommodate the
$12$ qubit quantum system. To define and code
the $12$ qubits through the eigenfunctions, let $\bz=(z_1,\ldots,
z_{12})^\dagger$ with $z_j=0$ or $1$, and
$\vec{k}=(k_1,\ldots, k_6)^\dagger=(z_1,\ldots,z_{6})^\dagger+ 2
(z_7,\ldots, z_{12})^\dagger$,
where~$\dagger$ denotes the transpose of a vector. The coordinates of
$\vec{k}$ take four integer values from
$0$ to $3$. We identify qubit state $|z_1\cdots z_{12}\rangle$ with
eigenfunction~$h_{\vec{k}}(\mathbf{x})$.
The quantum system is governed by Hamiltonian $H$ and evolves in time~%
inter\-val~$[0,1]$.
Let $V_\delta=e^{-i H \delta}$. We illustrate the quantum simulation
by approximating $V_\delta$ with
\[
U_\delta= \prod_{j=1}^6 [ e^{-i \bolds{\xi}^2_j \ell
\delta/4} e^{i \nabla_j^2 \ell\delta/2}
e^{-i \bolds{\xi}^2_j \ell\delta/4} ].
\]
Assume that the quantum system has an initial state at $t_0=0$:
\[
|\varphi_0 \rangle= \frac{1}{64}\sum_{z_j=0}^1 |z_1\cdots z_{12}
\rangle= \frac{1}{64} \sum_{k_j=0}^3
h_{k_1}(x_1) \cdots h_{k_6}(x_6),
\]
and final true state $|\varphi_m \rangle$ at $t_m=1$, where for
$j=1,\ldots, m$,
%e^{-i (z_1+\cdots+z_6+2 z_7+
% \cdots+ 2 z_{12}+3)}
%|z_1\cdots z_{12} \rangle= \frac{1}{64} \sum_{k_j=0}^3 e^{-i (k_1 +
% \cdots h_{k_6}(x_6) \]
%
\begin{eqnarray*}
|\varphi_j \rangle&=& e^{-i H t_j} |\varphi_0\rangle= V_\delta^j
|\varphi_0\rangle= \frac{1}{64}
\sum_{z_j=0}^1 e^{-i (z_1+\cdots+z_6+2 z_7+\cdots+ 2 z_{12}+3) t_j}
|z_1\cdots z_{12} \rangle\\
&=& \frac{1}{64} \sum_{k_j=0}^3 e^{-i (k_1 +\cdots+ k_6+3) t_j} h_{k_1}(x_1)
\cdots h_{k_6}(x_6).
\end{eqnarray*}
The approximation states in the quantum simulation are
\[
| \tilde{\varphi}_j\rangle= U_\delta^j |\varphi_0\rangle,\qquad
j=1,\ldots, m.
\]
Consider a path-dependent observable
\[
\bX= \frac{1}{20} \sum_{z_j=0}^1 (z_1+\cdots+ z_6 + 2 z_7+ \cdots+
2 z_{12}) \bQ_{ e^{-i H t_{z(b)}} |\bz\rangle},
\]
where $\bQ$ is a projection operator,
$\bz=(z_1,\ldots,z_{12})^\dagger$, $z(b)= \sum_{j=1}^{12} z_j
2^{j-1}$ corresponds to the number with binary
representation $z_1\cdots z_{12}$, and
\begin{eqnarray*}
e^{-i H t_{z(b)}} |\bz\rangle &=& e^{-i (z_1+\cdots+z_6+2 z_7+\cdots+2
z_{12}+3) t_{z(b)}}
|z_1\cdots z_{12} \rangle\\
&=& e^{-i (k_1 +\cdots+ k_6+3) t_{z(b)}} h_{k_1}(x_1)
\cdots h_{k_6}(x_6).
\end{eqnarray*}
We compute $\operatorname{tr}(\bX\rho)$ and $\operatorname{tr}(\bX^2 \rho)$ as follows:
\begin{eqnarray*}
\bX|\varphi_m\rangle&=& \frac{1}{20} \sum_{z_j=0}^1 (z_1+\cdots+
z_6 + 2 z_7+ \cdots+ 2 z_{12})\\
&&\hphantom{\frac{1}{20} \sum_{z_j=0}^1}
{}\times e^{-i (z_1+\cdots+z_6+2 z_7+\cdots+2 z_{12}+3)} |z_1\cdots z_{12}
\rangle\\
& =& \frac{1}{20} \sum_{k_j=0}^3 (k_1 +\cdots+ k_6+3)
e^{-i (k_1 +\cdots+ k_6+3)} h_{k_1}(x_1) \cdots h_{k_6}(x_6), \\
 \theta&=& \operatorname{tr}(\bX \rho) = \langle\varphi_m |\bX|\varphi_m\rangle
= \frac{1}{ 20 \times2^{12}} \sum_{k_j=0}^3 (k_1 +\cdots+ k_6+3) =
0.6, \\
%&& = \frac{1}{20 \times2^{12}} (\sum_{k_j=0}^3 6 k_1 \times4^5+ 3
% = \frac{1}{20 \times2^{12}} (6 \times6 \times4^5 + 3 \times4^6) =
\operatorname{tr}(\bX^2 \rho) &=& \langle\varphi_m |\bX^2 |\varphi_m\rangle=
\frac{1}{20^2 \times2^{12}} \sum_{k_j=0}^3 (k_1 +\cdots+ k_6+3)^2 =
0.37875, \\
% =\frac{1}{20^2 \times2^{12}} \sum_{k_j=0}^3 (6 k_1^2 + 30 k_1 k_2 +
%6^2 k_1 + 9) \\
%&& = \frac{1}{20^2 \times2^{12}} (6 \times14 \times4^5 +
% 30 \times6^2 \times4^4 + 6^3\times4^5 + 9 \times4^6) =
\operatorname{Var}(\hat{\theta}) &=& \frac{1}{n} [ \operatorname{tr}(\bX^2 \rho) - \theta^2 ]
= \frac{0.01875}{n}.
% \frac{1}{400 n} (151.5 - 12^2) = \frac{7.5}{400 n}=\frac{0.01875}{n}.
\end{eqnarray*}
Hence, we obtain the following expression for the MSE of $\hat{\theta}$:
\[
\mathit{MSE} = \frac{0.01875}{n} + (\langle\tilde{\varphi}_m |\bX|\tilde
{\varphi}_m \rangle- 0.6)^2.
\]
We need to numerically compute $\langle\tilde{\varphi}_m |\bX
|\tilde{\varphi}_m \rangle$
for the MSE evaluation. As~$H_\ell$ and $H_j$ are commutable,
$e^{-i H j \delta} = e^{-i H_1 j \delta} \cdots e^{-i H_6 j,\delta
}$, and
\[
e^{-i H j \delta} h_{k_1}(x_1) \cdots h_{k_6}(x_6) = \prod_{\ell=1}^6
e^{-i H_\ell j \delta} h_{k_\ell}(x_\ell).
\]
The numerical method in Zalka (\citeyear{Z1998}) can be used to evaluate $e^{-i
H_\ell j \delta} h_{k_\ell}(x_\ell)$
by repeatedly applying %$U_\delta$ as follows,
\[
U_\delta h_{k_1}(x_1) \cdots h_{k_6}(x_6) = \prod_{\ell=1}^6 [ e^{-i
\bolds{\xi}^2_\ell\delta/4}
e^{i \nabla_\ell^2 \delta/2}
e^{-i \bolds{\xi}^2_\ell\delta/4} h_{k_\ell}(x_\ell)
].
\]
We approximate $(\langle\tilde{\varphi}_m |\bX|\tilde{\varphi}_m
\rangle- 0.6)^2$
%compute $\langle\tilde{\varphi}_m |\bX|\tilde{\varphi}_m \rangle$
for $N=5000$ and $\delta$ ranging from $0$ to $0.01$ and then evaluate
MSE. The resulting MSE as a
function of $\delta$
decreases for $\delta$ from $0$ to $0.0035$ and then starts to
increase. Its unique minimum is achieved at
$\delta=0.0035$, which corresponds to $m=277$ and $n= 18$. Thus,
with total $5000$ times of state approximations allowed in the Monte
Carlo quantum simulation for
estimating $\theta$, the Monte Carlo strategy to minimize the MSE of
$\hat{\theta}$ is to
take $\delta= 0.0035$ in the quantum simulation scheme and repeatedly
simulate the quantum
system $18$ times.

\section*{Acknowledgments}
%Yazhen Wang is Professor, Department of Statistics, University of
%Wisconsin-Madison, Madison, WI 53706, USA. Email: yzwang@stat.wisc.edu
The author thanks editor Samuel Kou, the Associate Editor and an anonymous referee
for helpful comments and suggestions.

%suskaldyti doi

\printaddresses


\begin{thebibliography}{99}

%b1 ###
\bibitem[\protect\citeauthoryear{}{1997}]{AL1997}
\textsc{Abrams}, D. S. and \textsc{Lloyd}, S. (1997). Simulation of
many-body Fermi systems on a~quantum
computer. \textit{Phys. Rev. Lett.} \textbf{79} 2586--2589.

%%b2 ###
%quantum state generation and
%statistical zero knowledge. In \textit{Proc. 35th Annual ACP Symp. on
%Theory of
%Computing} 20--29. ACM, New York.

%b3 ###
%(2005). An invitation to
%quantum tomography. J. Roy. Statist. Soc. {\bf67}, 109--134.
\bibitem[\protect\citeauthoryear{}{2006}]{AS2006}
\textsc{Altland}, A. and \textsc{Simons}, B. (2006). \textit
{Interaction Effects in the Tight-Binding System.
Condensed Matter Field Theory}. Cambridge Univ. Press.
%b4 ###
\bibitem[\protect\citeauthoryear{}{2005}]{Aetal2005}
\textsc{Aspuru-Guzik}, A., \textsc{Dutoi}, A. D., \textsc{Love}, P.
J. and \textsc{Head-Gordon}, M. (2005).
Simulated quantum computation of molecular energies. \textit{Science}
\textbf{309} 1704.

%b5 ###
\bibitem[\protect\citeauthoryear{}{2003}]{BGP2003}
\textsc{Barndorff-Nielsen}, O. E., \textsc{Gill}, R. and \textsc{Jupp}, P. E. (2003). On quantum
statistical inference (with discussion). \textit{J. Roy. Statist. Soc.
Ser. B} \textbf{65}
775--816.
\MR{2017871}

%Berry, D. W., Ahokas, G., Cleve, R. and Sanders, B. C. (2006).
%Efficient quantum algorithms for simulating sparse Hamiltonians.
%ar.iv:quant-ph/0508139 [quant-ph].

%b6 ###
\bibitem[\protect\citeauthoryear{}{1998}]{BT1998}
\textsc{Boghosian}, B. M. and \textsc{Taylor}, W. (1998). Simulating
quantum mechanics on a quantum computer.
\textit{Phys. D} \textbf{120} 30--42.
\MR{1679863}

%L. (2007). Minimax and adaptive
%estimation of the Wigner function in quantum homodyne tomography with
%noisy
%data. Ann. Statist. {\bf35}, 465--494.

%and discrete-time
%quantum walk. ar.iv:0810.0312v2 [quant-ph].

%b7 ###
%doi:10.1038/nature07128. ISSN 0028--0836. PMID 18563154.
%http://www.nature.com/nature/journal/v453/n7198/full/nature07128.html.
\bibitem[\protect\citeauthoryear{}{2008}]{CW2008}
\textsc{Clarke}, J. and \textsc{Wilhelm}, F. (2008). Superconducting
quantum bits.
\textit{Nature} \textbf{453} 1031--1042.

%b8 ###
\bibitem[\protect\citeauthoryear{}{1985}]{D1985}
\textsc{Deutsch}, D. (1985). Quantum theory, the Church--Turing
principle and the
universal quantum computer. \textit{Proc. Roy. Soc. London Ser. A}
\textbf{400} 97--117.
\MR{0801665}

%b9 ###
%doi:10.1038/nature08121. ISSN 0028--0836. PMID 19561592.
%http://www.nature.com/nature/journal/vaop/ncurrent/pdf/nature08121.pdf.
%Retrieved 2009--07-02.
\bibitem[\protect\citeauthoryear{}{2009}]{Detal2009}
\textsc{DiCarlo}, L., \textsc{Chow}, J. M., \textsc{Gambetta}, J.
M., \textsc{Bishop}, L. S., \textsc{Johnson}, B.~R.,
\textsc{Johnson}, B.~R., \textsc{Schuster}, D.I., \textsc{Majer},
J., \textsc{Blais}, A., \textsc{Frunzio}, L.,
\textsc{Girvin}, S. M. and \textsc{Schoelkopf}, R.~J. (2009).
Demonstration of two-qubit
algorithms with a superconducting quantum processor. \textit{Nature}
\textbf{460}
240--244.

%b10 ###
%doi:10.1126/science.270.5234.255.
\bibitem[\protect\citeauthoryear{}{1995}]{D1995}
\textsc{DiVincenzo}, D. P. (1995). Quantum computation. \textit
{Science} \textbf{270} 255--261.
\MR{1355956}

%b11 ###
\bibitem[\protect\citeauthoryear{}{2005}]{D2005}
\textsc{Dziarmaga}, J. (2005). Dynamics of a quantum phase transition: Exact
solution of the quantum Ising model. \textit{Phys. Rev. Lett.} \textbf{95}
245701.

%%b12 ###
%(2002). Simulation of topological
%field theories by quantum computers. \textit{Comm. Math. Phys.}
%587--603.

%b13 ###
%doi:10.1007/BF02650179.
%http://www.springerlink.com/content/t2x8115127841630. Retrieved
%2007--10-19.
\bibitem[\protect\citeauthoryear{}{1982}]{F1982}
\textsc{Feynman}, R. P. (1982). Simulating physics with computers.
\textit{Int. J. Theor. Phys.} \textbf{21} 467--488.
\MR{0658311}

%%b14 ###
%needle in a
%haystack. \textit{Phys. Rev. Lett.} \textbf{79} 325--328.

%b15 ###
\bibitem[\protect\citeauthoryear{}{1982}]{H1982}
\textsc{Holevo}, A. S. (1982). \textit{Probabilistic and Statistical
Aspects of Quantum Theory}.
North-Holland, Amsterdam.
\MR{0681693}

%Statistics. New Jersey, Wiley. Second Edition.

%b16 ###
\bibitem[\protect\citeauthoryear{}{1978}]{K1978}
\textsc{Kato}, T. (1978). Trotter's product formula for an arbitrary
pair of self-adjoint
contraction semigroups. In \textit{Topics in Functional Analysis
(Essays Dedicated to M. G. Krein on
the Occasion of His 70th Birthday)}. \textit{Adv. in Math. Suppl.
Stud.} \textbf{3} 185--195. Academic
Press, Boston, MA.
\MR{0538020}

%b17 ###
\bibitem[\protect\citeauthoryear{}{2009}]{K2009}
\textsc{Kou}, S. (2009). A selective view of stochastic inference and
modeling problems in
nanoscale biophysics. \textit{Sci. China A} \textbf{52} 1181--1211.
\MR{2520569}

%b18 ###
%doi:10.1126/science.273.5278.1073.
%http://www.sciencemag.org/cgi/content/abstract/273/5278/1073.
%Retrieved 2009--07-08.
\bibitem[\protect\citeauthoryear{}{1996}]{L1996}
\textsc{Lloyd}, S. (1996). Universal quantum simulators. \textit
{Science} \textbf{273} 1073--1078.
\MR{1407944}

%Monroe, Don, Anyons (2008). The breakthrough quantum computing needs.
%New Scientist, 1 October issue.

%%b19 ###
%%doi:10.1126/science.1157233. PMID 18535240.
%%http://www.sciencemag.org/cgi/content/abstract/320/5881/1326.
%Multipartite entanglement among single
%spins in diamond. \textit{Science} \textbf{320} 1326--1329.

%b20 ###
%ISBN 0--521-63503--9. OCLC 174527496.
\bibitem[\protect\citeauthoryear{}{2000}]{NC2000}
\textsc{Nielsen}, M. and \textsc{Chuang}, I. (2000). \textit{Quantum
Computation and Quantum
Information}. Cambridge Univ. Press, Cambridge.
\MR{1796805}

%b21 ###
\bibitem[\protect\citeauthoryear{}{1995}]{S1995}
\textsc{Sakurai}, J. J. (1995). \textit{Modern Quantum Mechanics}.
Addison-Wesley, Reading, MA.
%
%%b22 ###
%(2003). Quantum random-walk search
%algorithm. \textit{Phys. Rev. A} \textbf{67} 052307.

%b23 ###
\bibitem[\protect\citeauthoryear{}{1994}]{S1994}
\textsc{Shor}, P. W. (1994). Algorithms for quantum computation:
Discrete logarithms
and factoring. In \textit{Proc. 35th Symp. on Foundations of Computer
Science} 124--134. IEEE Comput. Soc. Press, Los Alamitos, CA.
\MR{1489242}

%b24 ###
\bibitem[\protect\citeauthoryear{}{1999}]{SS1999}
\textsc{Sornborger}, A. T. and \textsc{Stewart}, E. D. (1999). Higher
order methods for simulations on
quantum computers. \textit{Phys. Rev. Lett.} \textbf{60} 1956--1965.

%problem of equilibration and
%the computation of correlation functions on a quantum computer.
%Preprint.

%b25 ###
\bibitem[\protect\citeauthoryear{}{1959}]{T1959}
\textsc{Trotter}, H. F. (1959). On the product of semi-groups of
operators. \textit{Proc.
Amer. Math. Soc.} \textbf{10} 545--551.
\MR{0108732}

%b26 ###
\bibitem[\protect\citeauthoryear{}{2007}]{W2007}
\textsc{Waldner}, J. B. (2007). \textit{Nanocomputers and Swarm Intelligence}.
ISTE, London.
%
%%b27 ###
%{Encyclopaedia of Mathematics} (M. Hazewinkel, ed.).
%Kluwer Academic Publishers, Dordrecht.

%b28 ###
\bibitem[\protect\citeauthoryear{}{1998}]{Z1998}
\textsc{Zalka}, C. (1998). Simulating a quantum systems on a quantum
computer. \textit{Proc. Roy. Soc. London Ser.~A} \textbf{454} 313--322.

\end{thebibliography}
\end{document}